# Crushing Modes of Aluminium Tubes under Axial Compression

**Florent Pled[1], Wenyi Yan[2] and Cui'e Wen[3]**

[1]Département de Génie Mécanique, Ecole Normale Supérieure de Cachan, France
[2]Department of Mechanical Engineering, Monash University, Clayton, Victoria 3800, Australia
[3]Center for Material and Fibre Innovation, Deakin University, Geelong, Victoria 3217, Australia

*Abstract:* A numerical study of the crushing of circular aluminium tubes with and without aluminium foam fillers has been carried out to investigate their buckling behaviours under axial compression. A crushing mode classification chart has been established for empty tubes. The influence of boundary conditions on crushing mode has also been investigated. The effect of foam filler on the crushing mode of tubes filled with foam was then examined. The predicted results would assist the design of crashworthy tube components with the preferred crushing mode with the maximum energy absorption.

## 1 Introduction

In the design of metallic energy dissipating systems, thin-walled circular aluminium tubes under axial loading conditions have been identified for several decades as very efficient impact energy absorbing elements, because of their progressive axial folding [1, 2]. When a circular thin-walled tube is crushed axially, it collapses either axisymmetrically or non-symmetrically, depending primarily on the ratio of the diameter to the thickness ($D/t$), but also on the ratio of the length to the diameter ($L/D$). The axisymmetrical mode is often known as the ring mode or concertina mode, while the non-symmetrical mode is referred to as diamond mode. The diamond mode is characterised by the number of lobes. For certain values of $D/t$ ratio, a tube may start collapse with the ring mode and then switch to the diamond mode, hence exhibiting a mixed mode. Tubes can also collapse in Euler-type buckling mode, which is a catastrophic collapse involving large bending of the tube and resulting in considerable loss of energy absorbing capacity. If the crushing modes of circular tubes with given geometric dimensions can be accurately predicted, then initiation of the preferred concertina mode with the highest energy absorption values can be guaranteed without the need for physical testing [3].

Recently, there has been an increasing interest in using aluminium foams as cores inside thin-walled metallic tubes for energy absorption. Previous study indicates that the energy absorption of the foam-filled tubes can be improved significantly compared to empty tubes. Therefore aluminium foam-filled tubular structures offer significant potential for industry applications where crashworthiness and weight are critical due to their unique combination of physical and mechanical properties, and excellent energy absorption.

Before quantifying the energy absorption by an empty tube or a foam-filled tube, it is important to understand the crushing modes as different crushing mode will lead to different energy absorption capacity. Finite element analyses are carried out in this paper to address this issue.

## 2 Numerical Model

Finite Element models were developed using the commercial package ABAQUS to simulate the crushing. A typical FE model consists of a cylinder clamped by two rigid compression plates. Both 4-node shell and 8-node solid elements are considered to model the empty tube. The foam core was modelled by using 8-node solid elements when modelling foam-filled tube.

The axial compression is simulated by moving down the upper plate while the lower plate is stationary. Several boundary conditions reflecting the connection between the empty tube and the plates are studied. To our knowledge, the influence of the boundary conditions on crushing mode of empty tubes has not been paid much attention in literature yet. All the boundary conditions studied in the crushing simulations are listed in Table 1.

Table 1 All the boundary conditions involved in the crushing simulation

| | Bottom of the tube | Top of the tube |
|---|---|---|
| $1^{st}$ boundary condition | tie constraint | roller |
| $2^{nd}$ boundary condition | Fixed $\begin{cases} u_r = 0 \\ u_\theta = 0 \end{cases}$ | fixed $\begin{cases} u_r = 0 \\ u_\theta = 0 \end{cases}$ |
| $3^{rd}$ boundary condition | Fixed $\begin{cases} u_r = 0 \\ u_\theta = 0 \end{cases}$ | roller |
| $4^{th}$ boundary condition | Roller | roller |

The value of the friction coefficient $f$ between the tube and the foam is chosen as 0.1. The materials data are obtained from uniaxial compression tests [4]. ABAQUS contains an extensive library to model the behaviour of various engineering materials. It includes models for metal plasticity and crushable foam plasticity.

The tube is an annealed aluminium Alloy A6063T5. Its Young's modulus $E_t$ and Poisson's ratio $\nu_t$ are 68.9 GPa and 0.33, respectively. The initial yield stress $\sigma_{Yt}$ is 48 MPa. The density of the aluminium tube $\rho_t$ is 2700 kg/m$^3$. The tube material is assumed to be elastic-plastic with isotropic strain hardening, which is described by a power-law work hardening. The hardening data were calibrated from uniaxial compressive tests curves [4].

The foam is treated as an isotropic homogeneous material. The relative density of the foam, defined as the fraction of space occupied by the solid, is determined by the porosity $P_r$ [3]:

$$\bar{\rho} = \frac{\rho_f}{\rho_a} = (1 - P_r), \tag{1}$$

where $\rho_f$ and $\rho_a$ are the densities of the foam and the aluminium solid, respectively. In the past, various analytical, experimental and numerical investigations have been carried out to derive material models for metal foams. It was found that the mechanical behaviour of metal foams depends on the relative density $\bar{\rho}$ significantly. The Young's modulus of the foam, $E_f$, can be estimated from the relative density $\bar{\rho}$ by [5]:

$$\frac{E_f}{E_a} = C_1 \cdot \bar{\rho}^2 \tag{2}$$

where $E_a$ is the Young's modulus of the aluminium solid. In the previous experimental study [4], the aluminium foam samples have closed-cell structure. $C_1$ is a constant related to the cell geometry of the foam, which was calibrated as 0.3 from the experimental results [4].

The solid cell wall material of aluminium have a Young's modulus $E_a$ of 68.9 GPa, a yield stress $\sigma_{Ya}$ of 170 MPa and a density of 2700 kg/m$^3$, according to the experimental data. The foam sample has a porosity $P_r$ of 91.5%, hence we can estimate the relative density $\bar{\rho}$ which is equal to 0.085 and the density of the foam, $\rho_f$, which is equal to 229.5 kg/m$^3$. According to Eq. (2), the Young's modulus of the foam, $E_f$, is obtained as 159 MPa. The linear elastic regime in the stress-strain curve of the foam sample is characterized by an elastic Young's modulus of 160 Mpa. A value of $\nu_f$ = 0.33 has been chosen in our numerical simulation [6].

The crushable foam plasticity model developed by Deshpande and Fleck [7] is applied to simulate the plastic behaviour of the aluminium foam. The yield function for this model is defined as:

$$\sqrt{\frac{\sigma_{eq}^2 + \alpha^2 \sigma_m^2}{1 + (\alpha/3)^2}} - Y = 0 \qquad (3)$$

where $\sigma_{eq}$ is the Von Mises effective stress, $\sigma_m$ is the mean stress or pressure stress, $\alpha$ is the shape factor of the yield surface and $Y$ is the hardening material function. The value of $\alpha$ can be determined by the uniaxial and hydrostatic compressive test, which is chosen as 1.58. The plastic Poisson's ratio $\nu^p$ = 0.18.

## 3 Results and Discussion

### 3.1 Mode classification chart for empty tubes

The crushing mode of an empty tube strongly depends on the geometry of the structure. First, the numerical simulations are compared with the available experimental data provided in Ref. [3], and analyzed with respect to the plastic folding mode. The first type of boundary conditions of Table 1 has been applied as it is close to the experimental conditions.

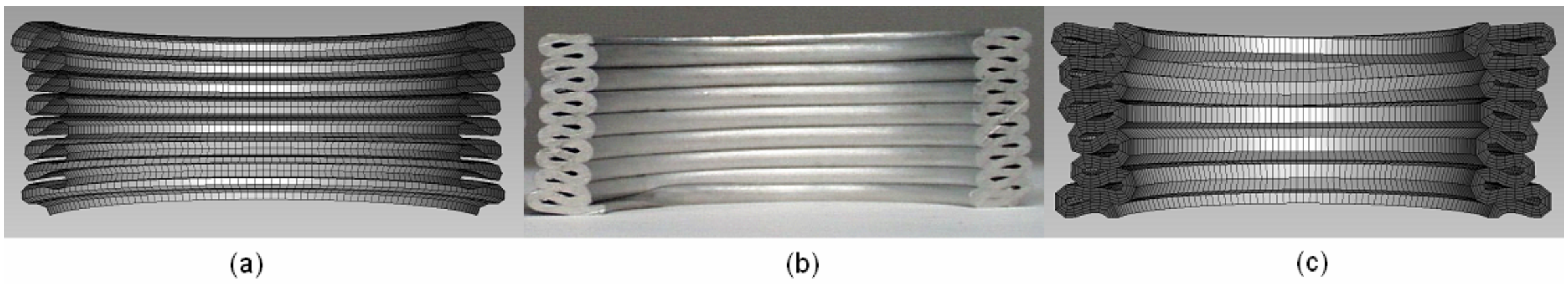

Fig. 1 Comparison of crushed shapes of empty tubes (a) shell-element simulation, (b) experiments and (c) solid-element simulation

The concertina modes have been correctly predicted. Fig. 1 shows the final crushed shape of an empty tube, compared with experimental observation. Both shell elements and solid elements were applied to model the tube and can successfully simulate the crushed shape. The shell element tube folds more frequently due to the ease with which the elements can rotate. The shell element model predicts 15 folds, which is exactly the number of folds observed in experiments, whereas the solid element model predicts only 12 folds.

Secondly, parametric study has been carried out to understand the influence of geometric parameters on the buckling modes of empty tubes. The range of geometries used ensured that various crush modes were observed. Two hundred and four tubes were considered with varying $L/D$ ratio (from 1 to 12) and $D/t$ ratio (from 10 to 500). Due to the small thickness of the tubes, shell elements were applied in the modelling. The crushed shapes have been successfully simulated for all the different crushing modes.

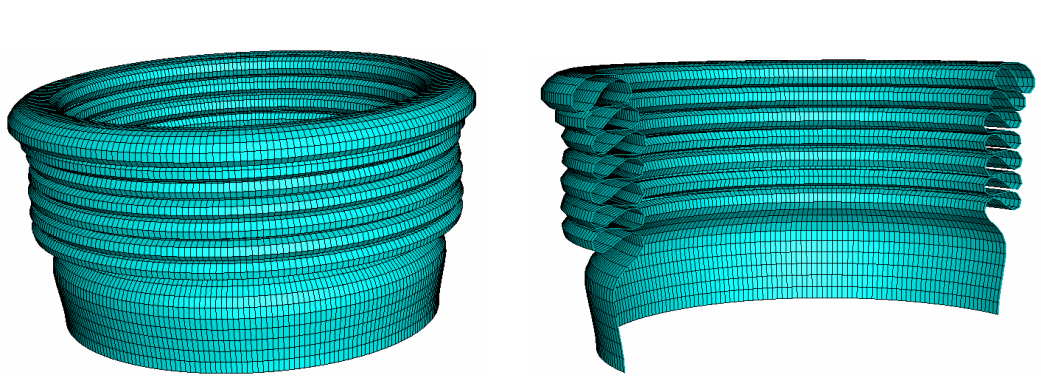

Fig. 2 Axisymmetric concertina mode and Cut view of concertina mode

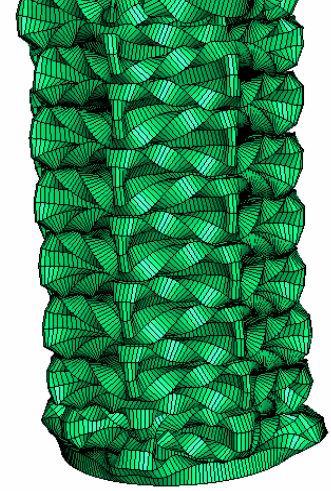

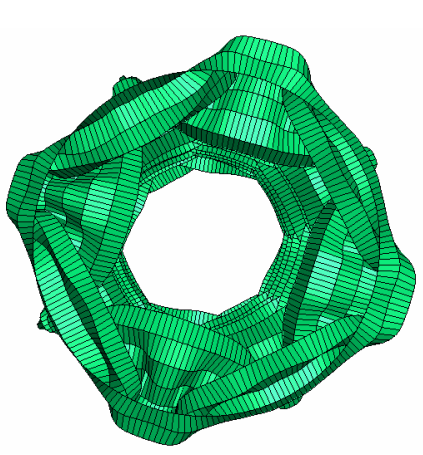

Fig. 3 Diamond 3 lobe mode and Bottom view of diamond mode

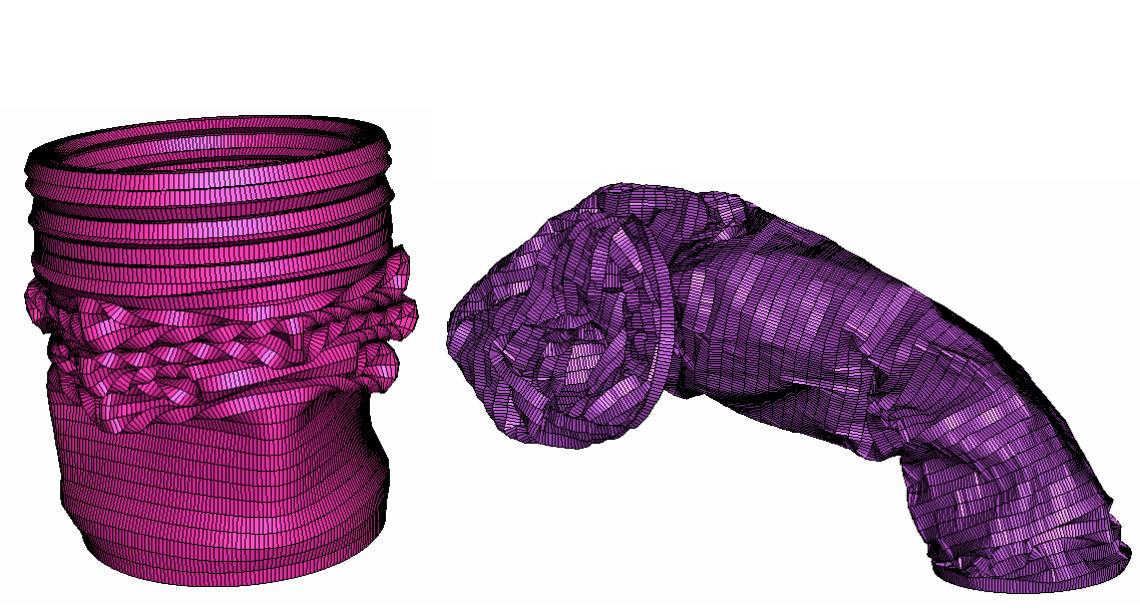

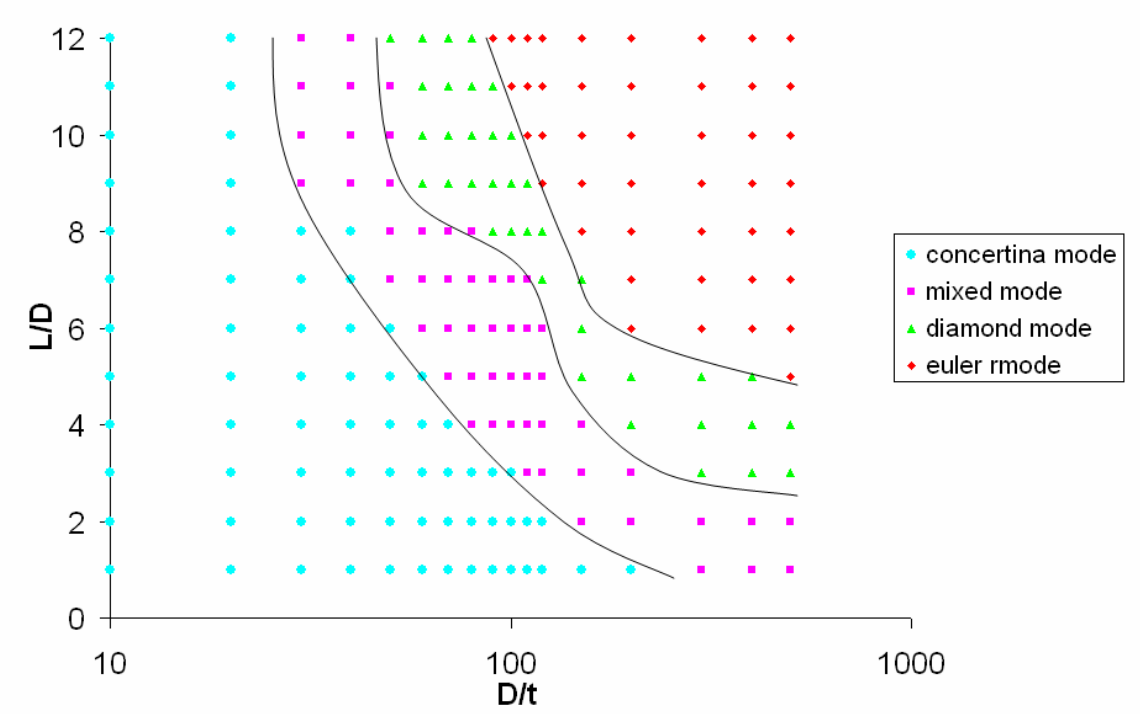

Fig. 4 Mixed mode and Euler-type buckling mode

Fig. 5 Mode classification chart for aluminium tubes

The results shown in Fig. 5 are consistent with the previous works. Our chart covers a sufficiently large range of $D/t$ and $L/D$. For short and thick tubes, concertina mode appears. When the thickness of the tube is reduced and the length of the tube is increased, mixed mode occurs and then diamond mode is present. For long and very thin tubes, Euler-type buckling mode takes place.

## 3.2 Influence of boundary conditions on buckling modes for empty tubes

Four empty tubes have been selected to investigate the influence of the boundary conditions on crushing modes. Under the first type of boundary conditions, four different crushing modes are expected from these four representative tubes, see Table 2.

Table 2 Dimension parameters of aluminium tubes

| Tube | Length $L$ (mm) | Outer Diameter $D$ (mm) | Wall thickness $t$ (mm) | $L/D$ | $D/t$ | Expected crush modes |
|---|---|---|---|---|---|---|
| 1 | 150 | 50 | 1.67 | 3 | 30 | Concertina |
| 2 | 200 | 25 | 0.28 | 8 | 90 | Diamond |
| 3 | 200 | 40 | 0.44 | 5 | 90 | Mixed |
| 4 | 210 | 30 | 0.06 | 7 | 500 | Euler |

For the tube 1, concertina (axisymmetric) mode was observed for all the boundary conditions in the simulation. Nevertheless for the two symmetric boundary conditions, (2) and (4), the folds appear at both sides of the tube, as shown in Fig. 6.

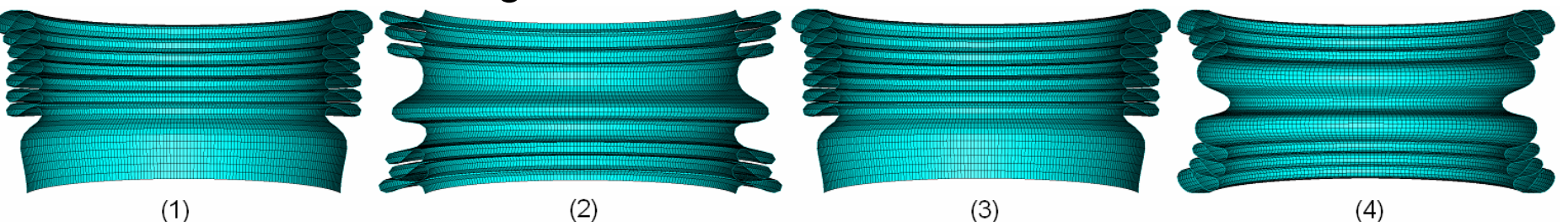

Fig. 6 Comparison of crushed shapes of Tube 1 for different boundary conditions (1) tie constraint-roller, (2) fixed-fixed (3) fixed-roller and (4) roller-roller

For the Tube 2, diamond mode were observed for the boundary conditions (1) and (4) in the simulation, whereas a mixed mode was observed for the boundary conditions (2) and (3), as shown in Fig. 7. A refinement of the mesh has been done to see the influence of boundary conditions for the Tube 2. We observed that mixed modes appeared for the boundary conditions (1), (3) and (4) and concertina mode appeared for the boundary condition (2). Therefore the boundary condition (2) seems to be the best for obtaining axisymmetric folds and absorbing a higher amount of energy.

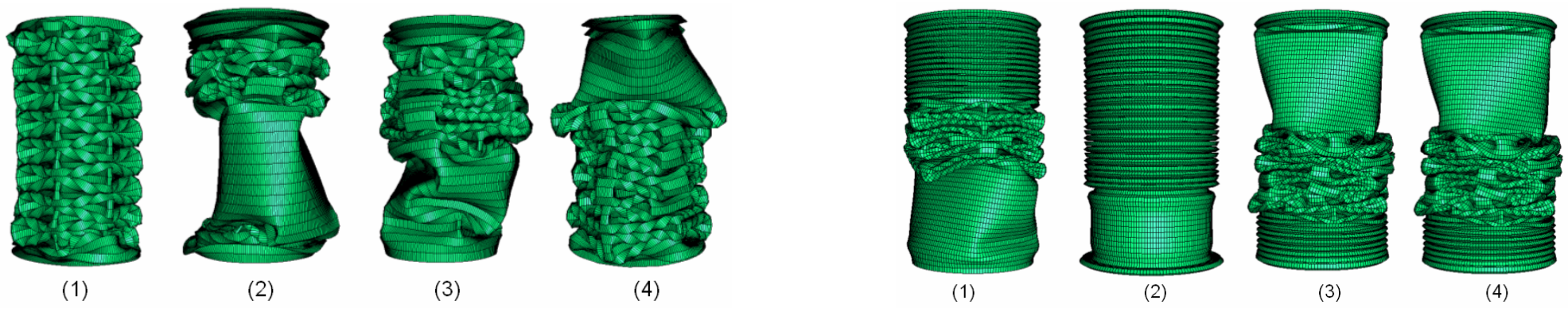

Fig. 7 Comparison of crushed shapes of Tube 2 for different boundary conditions (1) tie constraint-roller, (2) fixed-fixed (3) fixed-roller and (4) roller-roller

For the Tube 3, mixed mode was observed for boundary conditions (1), (2) and (4), but concertina mode was observed for the boundary condition (3) in the simulation, as shown in Fig. 8. Therefore when one end of the tube is fixed, a concertina mode appears instead of a mixed mode.

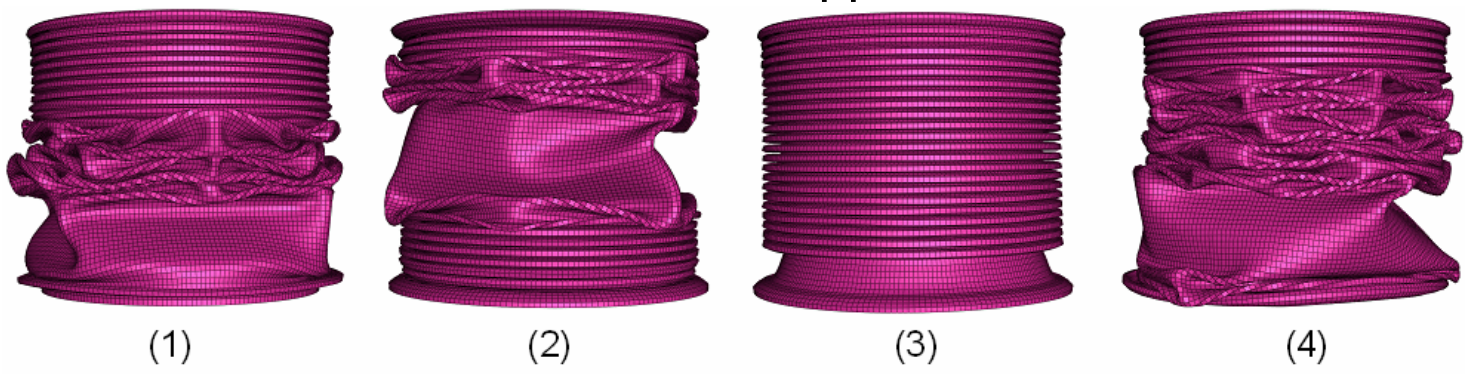

Fig. 8 Comparison of crushed shapes of Tube 3 for different boundary conditions (1) tie constraint-roller, (2) fixed-fixed (3) fixed-roller and (4) roller-roller

For the Tube 4, Euler buckling (catastrophic collapse) crush mode was observed for all boundary conditions in the simulation. When one end of the tube is fixed or tied to a plate, this part of the tube stay in contact with the plate and when a roller boundary condition is applied at a end of the tube, this end of the tube slides over the plate, as shown in Fig. 9.

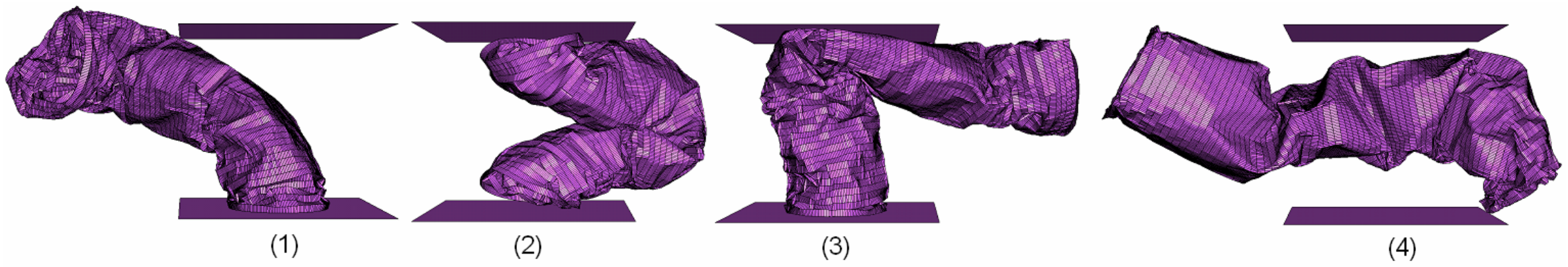

Fig. 9 Comparison of crushed shapes of Tube 4 for different boundary conditions (1) tie constraint-roller, (2) fixed-fixed (3) fixed-roller and (4) roller-roller

### 3.3 Influence of foam fillers on the crushing mode

According to experiments, an interaction effect usually exists between the tube wall and the foam core. The deformation behaviour of aluminium foams provides a resistance to the inwards deformation of the tube wall. Hence it causes an outwards expanding of the progressive buckling of tube wall, or a shift of deformation mode from diamond to concertina with foam-filling. Numerical simulations of the Tubes 1, 2 and 3 filled with aluminium foam have been carried out under the first set of boundary conditions. The numerical results carried out in this part are of great importance since they offer the possibility to predict the potential effects of filling on the buckling modes of a tube.

The effect of foam-filling for Tube 1 has two consequences: Foam-filling did not change the deformation mode, but it causes a decrease in the plastic fold length. Indeed the presence of aluminium foam results in shorter folding length in experiments [4]. Moreover, foam filling also caused the progressive buckling expanding outwards noticeably compared to the tube without foam, as shown in Fig. 10.

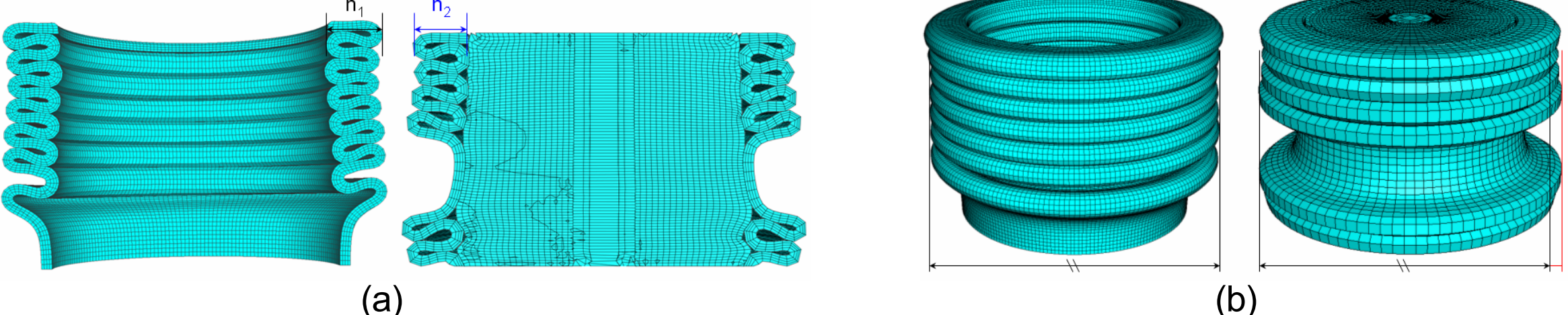

Fig. 10 Comparison of crushed shapes of the empty Tube 1 and the foam-filled tube, showing the same crushing mode and a decrease in the plastic fold length (a), but an outwards expansion of the buckling (b).

Foam-filling changed the deformation mode from diamond mode to Euler-type buckling mode and concertina mode for the Tube 2 and the Tube 3, respectively, see Fig. 11.

The numerical model seems to capture quite correctly the contribution of the foam for the concertina mode. Indeed, the modelled deformed shape of Tube 1 with foam matches the experimental deformed shape. The numerical results of Tube 3 show good agreement with the experimental observation, as the progressive buckling mode changed from non-symmetric mixed mode into concertina mode. The numerical results demonstrate the deformation shift of Tube 2 from diamond mode to Euler-type buckling mode due to foam filling. In the reported experimental work [4], foam filler of the tube converts the diamond mode into concertina mode, but the geometry of the tube is different from that of Tube 2. Furthermore the foam used is light (the porosity of the foam core was 91.5 %) compared to that used in the experiments (90 % of porosity). Therefore it could be interesting to change the foam properties and to see the influence on the crushing mode of diamond. The interaction effect between the tube and foam might be affected by not only the tube dimension parameters, but also the properties of the foam. As foam has been defined as an isotropic homogeneous material, only the effect of changing the density of foam has been investigated. The intention was to know if the shift from diamond mode to concertina mode due to foam filling mode could occur for a tube filled with a foam of higher density. The porosity of the foam has been changed from 91.5% to 88%. Finally the Tube 2 filled with a denser foam collapses in Euler-type buckling mode and not concertina mode. Further systematic work is required to completely understand the influence of foam filler on the crushing mode.

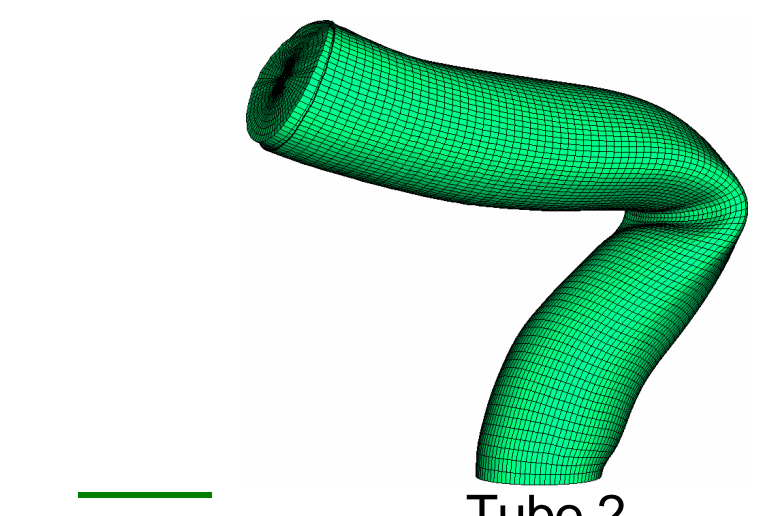

Tube 2

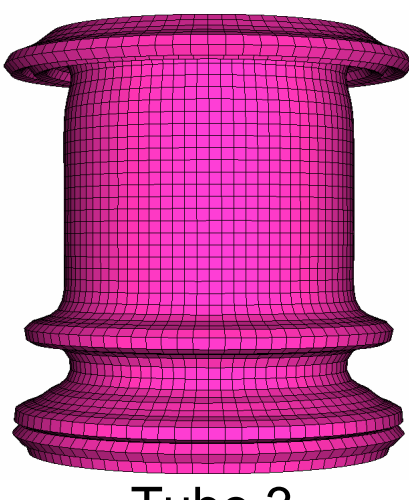

Tube 3

Fig. 11 Crushed shapes of the Tube 2 and the Tube 3 filled with a foam of 91.5 % of porosity, showing that progressive crushing mode changes from non-symmetric mode into Euler-type buckling mode and concertina mode for the Tube 2 and Tube 3, respectively

## 4 Conclusions

Numerical simulations of the crushing of aluminium tubes with and without foam filler have been carried out. The experimental observations of the post-buckling deformation shape have been successfully predicted. The influence of geometric parameters on buckling modes of empty tubes has been investigated and a mode classification chart has been established. It shows a quite good agreement with previous experimental charts. Different boundary conditions for the tube have been studied. It shows that when at least one end is fixed is the best set of boundary conditions for energy absorption because it creates more axisymmetric folds, whereas when the two ends of the tube are free is the worst set of boundary conditions for energy absorption. Our numerical observations give excellent results compared to the experiments for the concertina mode. Numerical results for the diamond mode show that Euler-type buckling mode could occur for foam-filled tube. Indeed progressive crushing mode did not change from diamond mode to axisymmetric concertina mode numerically, but aluminium foam converted the diamond mode into the Euler-type buckling mode, even with a foam of higher density. Therefore for thick and short tubes, foam-filling seems to be effective in enhancing the energy-absorption performance, whereas if a tube is too thin, it may have a strong tendency to undergo Euler-type buckling, greatly reducing the energy absorption.